# Dynamic Models of Learning and Education Measurement

Lei Bao

Department of Physics, The Ohio State University,
191 W Woodruff Ave., Columbus, OH 43210

**Abstract**

Pre-post testing is a commonly used method in physics education community for evaluating students' achievement and or the effectiveness of teaching through a specific period of instruction. A popular method to analyze pre-post testing results is the normalized gain first brought to the physics education community in wide use by R.R. Hake. This paper presents a measurement based probabilistic model of the dynamic process of learning that explains the experimentally observed features of the normalized gain. In Hake's study with thousands of students' pre-post testing results, he observed that on average 48 courses employing "interactive engagement" types of instruction achieved average normalized gains about two standard deviations greater than did 14 courses subjected to traditional instruction. For all courses the average normalized gains had a very low correlation +0.02 with average pretest scores. This feature of the normalized gain has allowed researchers to investigate the effectiveness of instruction using data collected from classes with widely different average pretest scores. However, the question of why the average normalized gain has this feature and to what extent this feature is generally present is not well understood. In addition, there have been debates as to what the normalized gain actually measures, and concerns that it lacks a probability framework that undergirds psychometric methods such as Item Response Theory (IRT). The present model leads to an explanation of the observed features of the normalized gain, connects to other models such as IRT, and shows that the normalized gain does have a probability framework but one different from that emphasized by IRT.

## I. Introduction

Quantitative education assessment can be designed to serve different goals. A standard goal is the evaluation of individuals' proficiency on specific skills or knowledge compared to others in the same population. This allows teachers to grade students into performance levels, but doesn't indicate how and when the knowledge is developed. Such assessment summarizes the outcome of all learning since day one. However, in education research, it is usually of more interest to assess the effectiveness of teaching and learning during a specific (often short) period of time such as the duration of a course. By controlling variables, this measurement can be designed to evaluate (a) the relative scale of learning ability of individual students among a population going through identical instruction or (b) the effectiveness of different instructional methods applied on a single population. This type of assessment emphasizes the measurement of students' changes (gains) on certain knowledge scales, which requires at least two measurements conducted at different times.

A popular method in physics education is to use pre- and post- tests to evaluate the relative effectiveness of instruction on student learning. The results from the two tests are evaluated with an index ($g$), the ratio between the score difference of post- and pre-test and the maximum possible value of that difference, i.e., $g = (y-x)/(1-x)$, where $x$ is the pretest score, $y$ is the posttest score, and scores are scaled into the region of 0~1. This half-century-old pre/post gain index was utilized independently by Hovland et al. [1a] who called it the 'effectiveness index,' Gery [1b] who called it the "gap-closing parameter," and Hake [1c,d] who called it the "normalized gain". For a class the *average* normalized gain can be calculated either by averaging the single-student gains or by using the average pre- and post-test scores to calculate the class-average-score gain. For a discussion of these two types of averages see Hake [1c] and Bao [17].

In his experiment with 62 courses and N = 6542 students [1c], Hake observed that, on average, 48 courses employing "interactive engagement" types of instruction achieved average normalized gains <g> about two-standard deviations greater than 14 courses subjected to "traditional" instruction. For all courses the average normalized gains had a very low correlation +0.02 with pretest scores. This suggested that normalized gain might be used by researchers to disentangle the effects of instruction from student backgrounds, and since Hake's survey many research groups have done so with apparent success. However, the question of why across different classes and populations the average normalized gain was found to be virtually uncorrelated with the average pretest score is not well understood. In this paper, a probabilistic theoretical model is introduced to explain the possible mechanisms underlying the normalized gain. Further, the model is expanded to explore the fundamental features of popular education assessment methods and models such as the Item Response Theory [2, 3], and to make connections among the different methods.

**II. The Measurement Representation of Score Based Assessment and a Dynamic Model of Learning**

The most important and also difficult part in modeling learning is to define and quantify students' behavioral states. For example, the definition of "knowledge" is highly dependent on the context in which the term is used. Here, an operational approach is used which defines knowledge as one's internal cognitive function that can produce specific types of test results under given context situations. The properties of one's knowledge (or cognitive function) cannot be directly probed but can be inferred through measurement. The measurement results are "collapsed" states of one's cognitive function under specific constraints and conditions, and therefore are context dependent.[3-7] The settings of a particular measurement determine the types of representation for a person's measured knowledge states. For example, if a score is used as the measurement variable, then the features of students' knowledge are represented in a score-representation. In research on education and learning, researchers also have establish many cognitively based representations and measurement approaches, in which student knowledge is modeled and represented with hypothetical mental constructs and processes, such as misconceptions, mental models, facets, p-prims, etc. [3,8-12]

Without applying a particular cognitive framework, a student's knowledge on a single concept topic can be generally described with a combination of three situations: 1) scientifically correct knowledge, 2) scientifically incorrect knowledge, and 3) lack of relevant knowledge. Depending on the choice of measurement representations, the three situations produce different measurement outcomes. In a score representation, the correct knowledge is measured with the probability for a student to produce correct answers. Thus, the 2nd and 3rd situations are lumped together as the student's inability to produce correct answers, which is reflected by the complementary part of the score (1-s). The measurement is then reduced to a single effective dimension. Apparently, the reduction of dimensions for measurement often increases uncertainties in the interpretation the measurement results and limits the inferences on the possible cognitive processes underlying students' behavior. However, the benefit of a score-based measurement is its simplicity and its independence from potential biases of cognitive models.

Then what can be represented by a measured score? Due to the complexity of cognitive processes, there are many possible educational and or cognitive variables underlying an observed score, most of which are difficult to be precisely determined. For example, a low score can be the result of a range of cognitive processes such as incorrect applications of correct knowledge, "correct" applications of incorrect knowledge, applications of irrelevant knowledge, or "unlucky" guesses. In this paper, these possible causes to a measured score will not be explicitly expressed in the model and the variations caused by these processes will be treated in general as uncertainties of random origins. Based on the simplification, student's knowledge state is then described with a score representation that has two states: measured-correct and measured-wrong. Obviously, the measured-wrong state doesn't have to be caused by incorrect knowledge but is just not measured as correct with a particular instrument in a particular measurement instance.

A measurement instrument (or a test) often consists of a finite set of questions with different context settings that probe various features of a student's knowledge. Therefore, a quantitative measure of student's knowledge is evaluated as pieces of measured knowledge, each of which can be in a state of measured-correct or measured-wrong. As discussed earlier, the measured-correct and measured-wrong states may be the result of a range of factors that might have occurred either randomly or systematically. For simplicity, the details of these processes are not addressed in this paper and it is generally assumed that (1) the measured-wrong state represents in large the application of scientifically incorrect knowledge and the lack of appropriate knowledge, and (2) the measured-correct state represents in large the successful application of scientifically correct knowledge. Details on modeling possible inference errors can be found in the reference [12].

For a particular test, one can identify a list of pieces of measured knowledge, which defines the "measurement window" of the instrument. Such a measurement window is a subset (usually a representative subset) of the knowledge domain of a particular course. Obviously, the ceiling and floor effects of a measurement are the natural results of the measurement window. Further discussion on this issue will be given in the discussion section. Usually, it is also assumed that students' performance in the measurement window can be used to predict their behavior in knowledge areas outside the measurement window with some statistical uncertainty; however, this assumption doesn't alter the model discussed in this study. Thus, a student's score can be interpreted as the quantity of the student's measured-correct knowledge inside the measurement window. Since the size of the measurement window is fixed for a particular test, the score space is conserved and the complementary part of the score (1−s) gives the quantity of the student's measured-wrong knowledge. Note that this measurement representation only has one effective dimension.

Using this two-state score representation, an individual student's measured knowledge states can be represented with a "two-level" system (see Figure 1). The C level (upper level) represents the measured-correct knowledge, the quantity of which is given by the score. The W level (lower level) represents the measured-wrong knowledge and its quantity is given by the complementary part of the score.

The education system or environment is simplified into two objects: (1) a group of identifiable students, and (2) the education environment that interacts with the individual students. The instruction in a course is treated as an integration of many teaching-learning activities, in which students can interact with the environment. To a student, each teaching-learning activity is a "learning incident", which may or may not result in observable changes of the student's internal properties. In the score representation, observable changes will appear as transitions between the two levels of a student's measured-knowledge states.

As a physics analogy, one can consider students as particles each having a unique two-level system. The instruction is a stream of incoming learning incidents that can impact the students and cause transitions in their two-level systems. When a student encounters a learning incident, possible transitions can be no-change, excitation, and decay. Therefore, for a group of students going through the same instruction, the differences among individuals' learning can be represented as different excitation and decay coefficients corresponding to the features of the transition processes.

It is well understood that students' existing knowledge can affect their learning both positively and negatively. [3,8,13] The fundamental idea of the model presented in this paper is to express the process of students' learning in terms of a set of differential equations with coefficients representing the possible relations between students' existing knowledge state and their measured outcome of learning. In the score representation, one can consider three general types of transitions of a student's measured knowledge (see list below) that each can result in three observable changes of a student's measured knowledge state: excitation, no-change, and decay. Here only transitions due to cognitive processes are considered. Measurement results due to non-cognitive stochastic processes such as guessing and general mistakes are not included and will be treated as random errors. Below, three typical types of transitions are discussed to exemplify the theoretical formalism.

1. $\alpha$-type transitions: Changes of measured-correct knowledge due to constructive and or destructive interactions between the instruction and the student's measured-wrong knowledge. This reflects a process of direct external impact on changing students' incorrect or missing knowledge. For example, the instruction may directly address content areas that students have difficulties (defined in terms of producing wrong answers). If the process is constructive, it will help students learn to produce correct answers on a test. It can be generally agreed that the traditional instruction often operates in this manner, i.e., students learn passively through memorizing the pieces provided by the instruction. A destructive process of this type can be interpreted as if the instruction enhances the students' incorrect knowledge, which later inflicts a larger barrier for students to learn the correct knowledge or even cause some unstable correct understandings to change back to incorrect ones.

2. $\beta$-type transitions: Changes of measured-correct knowledge due to constructive and or destructive interactions between the instruction and the student's measured-correct knowledge. This reflects a process of direct external impact on students' correct knowledge. For example, the instruction may directly address content areas that students already know. If the process is constructive, it enhances the students' correct knowledge, which can later help students learn new knowledge or facilitate students to change some of their unstable/transitional incorrect understandings to correction ones. A destructive process of this type is also possible. For example the instruction may confuse some students and cause them to revert to their incorrect knowledge.

3. $\gamma$-type transitions: This reflects a process with significant internal interactions, in which both the correct and the incorrect knowledge are activated – a typical associative process of learning. Such interactions can cause a change of measured-correct knowledge through constructive and or destructive interactions between the student's correct and incorrect knowledge. A constructive process of this type is often the goal of the interactive engagement learning environment, in which students use constructive approaches to develop new understanding and revise their incorrect understanding based on their existing knowledge. A destructive process is also possible. Since students' incorrect knowledge is often difficult to change, the interactions between correct and incorrect knowledge can lead to both favorable and unfavorable outcomes. For example, before achieving a generalized understanding students can vacillate between using correct and incorrect knowledge on questions that are equivalent but are designed with different context features [3,5].

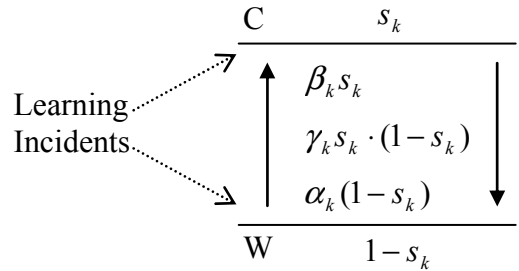

FIG. 1. Two-level systems of $k^{th}$ student's measured knowledge states. The measured-correct (C) knowledge is given by the student's score $s_k$, and the measured-wrong (W) knowledge is given by $(1-s_k)$.

When using pre- and post-test to measure students' learning gains, the result doesn't provide any information about intermediate states during the instruction. Therefore, the changes of a student's measured knowledge states are treated as having occurred randomly through the entire sequence of learning incidents.

In a general form, the probability for a student to have a change in the measured knowledge state is proportional to the product of the probability for instruction to activate/address relevant knowledge domain (the probability for a "hit") and the probability for that learning interaction to make impact on the student knowledge state (the effectiveness of a "hit"). For example, consider the $\alpha$-type transition, in which the instruction directly addresses students' incorrect knowledge. The probability for a student to have an increase or a decrease of his/her measured-correct knowledge in a arbitrary learning incident depends on the total quantity of the measured-wrong knowledge and the effectiveness of the learning incident to make impact on the student's wrong knowledge once activated. (This latter part is in theory expressed with a conditional probability discussed in Section VI.) That is, if a student has more incorrect knowledge, it is more likely for this student to encounter a learning incident that addresses some parts of his/her incorrect knowledge than a student with less incorrect knowledge. The transition rate can then be modeled in terms of the product of the quantity of the measured-wrong knowledge and the effectiveness of learning incidents.

Suppose one can perform multiple "identical" or "equivalent" measurements with students in a knowledge domain during the instruction. The result of such measurements for the $k^{th}$ student is represented with the student's score $s_k(t)$. Let $\alpha_k$ be the coefficient representing the effectiveness of the instruction in terms of α-type transitions. The learning rate of the α-type transition per learning incident can be represented by:

$$\frac{ds_k}{dt} = \alpha_k \cdot (1 - s_k). \tag{1}$$

Here, $t$ represents the sequence of learning incidents rather than the actual time. Eq. (1) represents the ensemble results of a large number of interactions with a continuous stream of learning incidents.

As an analogy, we can consider the quantity of a student's measured-wrong knowledge as the student's "cross-section" for impacts of the incoming learning incidents, whereas $\alpha_k$ gives the combined α-process excitation and decay coefficient of each impact (which can be interpreted as the teaching-learning impact coefficient). Here, the probability of a "hit" (or the cross-section for impact) is modeled as the result of a random process; therefore its magnitude is proportional to the quantity of student measured wrong knowledge. This assumes that instructors don't have additional means of obtaining information about students' incorrect knowledge. In practice, depending on features of specific education settings such as the use of formative assessment, immediate feedback, and tutoring, instructors may have access to additional information about students' incorrect knowledge, which will change the dynamics of the learning process and the formulism of the model. Taking the random hit assumption and combining all three processes (see Figure 1), the overall equation is obtained:

$$\frac{ds_k}{dt} = \alpha_k(1 - s_k) + \beta_k s_k + \gamma_k s_k(1 - s_k). \tag{2}$$

With this model, one can explore the dynamics of the different processes and their underlying assumptions to model the education measurement outcomes. When each of the processes is considered independently, the solutions are obtained below:

α-process: $s_k(t) = 1 - (1 - s_{k0})e^{-\alpha_k t}$ (3)

$\beta$-process: $s_k(t) = s_{k0} e^{\beta_k t}$                                                      (4)

$\gamma$-process: $s_k(t) = \dfrac{1}{1 + [(1 - s_{k0})/s_{k0}] \cdot e^{-\gamma_k t}}$                         (5)

Here, $s_{k0}$ represents the measured score of student $k$ at $t = 0$, which is usually the pretest score. Figure 2 shows the general shapes of the $s_k(t)$ curves of Eqs. (3, 4, 5) when all the coefficients are positive: (a) the $\alpha$-process exponentially decreases the gap between the maximum score (=1) and the student's score; (b) the $\beta$-process on the other hand exponentially increases the student's score unbounded; (c) the $\gamma$-process shows a logistic "S" shape which is widely used in modeling transition processes in neural science, social dynamics, and education assessment (e.g., the methods in Rasch Model and Item Response Theory).

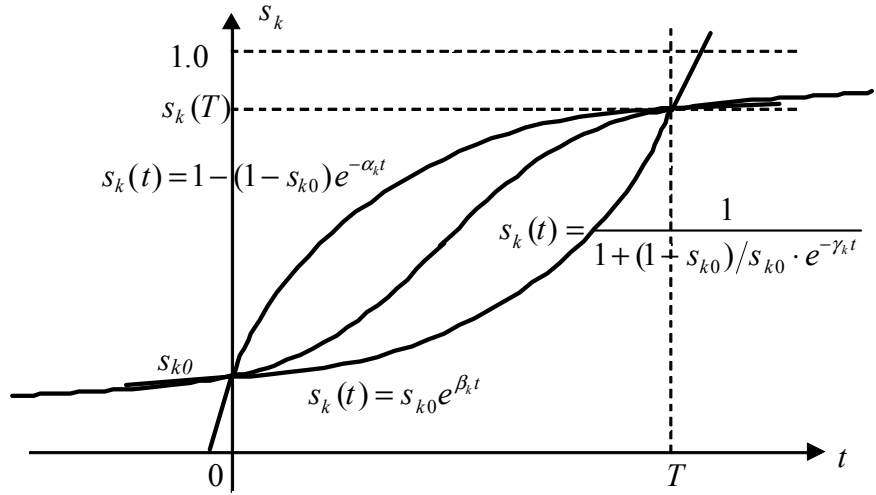

FIG. 2. Score $s_k$ as a function of time (learning incidents) for $\alpha$, $\beta$, and $\gamma$ processes as discussed in the text. The score varies from the pretest score $s_{k0}$ at $t = 0$ to the posttest score $s_k(T)$ at $t = T$.

It is worth clarifying that the model presented here is a measurement based probabilistic model, which describes the probabilistic relation between students' measured knowledge states and the measured changes of such states. It doesn't model any cognitive processes of how the learning occurs. Therefore, this model is fundamentally different from a cognitive model that describes the actual processes of learning. The coefficients defined in the three dynamic processes are used to parameterize the probabilities of measured changes in scores due to the combined effects of a wide variety of possible cognitive processes of learning. In this study, the possible substructures of the coefficients are not explored.

In current education settings, it is reasonable to assumed that the $\beta$-process, which represents the situation that students will spontaneously develop new knowledge based on their existing correct knowledge, has a very small success rate. As a result, this process is not analyzed in detail in this paper. In addition, the combined effects of learning throughout a course period often show a positive gain on measured knowledge; therefore, the corresponding decay forms of the processes (in which the coefficients are negative) will not be discussed in this paper.

**III. Model of Learning under Dominant External Influence**

To model the learning dynamics, one may first consider two general education settings: controlled and open. In a controlled format, the instruction controls what students learn and how they learn. This is a common feature of the existing lecture-based environment. The students often passively follow the instruction and rely heavily on memorizations [14]. In an open format, the instruction allows active learning activities, in which students construct new understanding by conducting self-controlled investigative learning of the new knowledge domain. Features of the open format are becoming the emphasis of recent curriculum development in physics education [13,15]. However, current strategies often target selected content areas to help students restructure their existing misconceptions so that they can develop correct understandings. There are also significant guidance and constraints on how and what students may explore during a learning session. To this end, students' learning is still heavily guided by the instruction. The changes to students' incorrect knowledge depend primarily on whether the instruction addresses the specific content areas that the students have difficulties with. Therefore, both the traditional instruction and the guided constructive methods have a significant to dominant portion of the $\alpha$-type processes. Obviously, the guided constructive approaches often have significantly higher probabilities for favorable transitions (larger $\alpha$ coefficients) than the traditional instruction does.

If it is assumed that the $\alpha$-process dominates, i.e., that both the $\beta$ and $\gamma$ processes can be neglected, then the solution of Eq. (1) is Eq. (3), rewritten in Eq. (6):

$$s_k(T) = 1 - (1 - s_{k0}) \cdot e^{-\alpha_k T} . \tag{6}$$

Here, $s_{k0}$ is the $k^{th}$ student's pretest score; $(1-s_{k0})$ represents the quantity of the initial measured-wrong knowledge; $T$ is the total duration of the instruction; and $s_k(T)$ is the posttest score of the $k^{th}$ student.

Notice that during instruction a single student can acquire knowledge at any learning incident that may be random in the time domain. The relation described in Eq. (3) represents the ensemble results of a large number of interactions with a continuous stream of learning incidents and should not be regarded as the exact pattern that a particular student would follow during a short period of instruction.

For a single student, we can calculate the normalized gain:

$$g_k = \frac{s_k(T) - s_k(0)}{1 - s_k(0)} = 1 - e^{-\alpha_k T}. \tag{7}$$

The result shows that the normalized gain is expressed in terms of the $\alpha$ coefficient and contains no explicit terms of the pretest score $s_{k0}$.

Therefore, *the normalized gain will not correlate with the pretest score if the α coefficient does not correlate with the pretest score*. On the other hand, the normalized gain is always positively correlated with the $\alpha$ coefficient. Since the $\alpha$ coefficient represents the effectiveness of the interactions between individual students and the learning incidents, it is related to both the effectiveness of instruction and students' general learning abilities. Eq. (7) provides a theoretical basis for using normalized gain to assess learning gains.

In general, the $\alpha$ coefficient can be related in various ways to students' initial knowledge states in a particular content domain. For example, consider two graduates from the same high school. Suppose they have similar learning ability as judged by similar high-school GPA and SAT scores, but in high school one student took an elective physics course while the other took a chemistry course instead. Both courses are elective and are not included in determining students' ranking/ability. On average, the general learning ability of the two students can be considered similar. However, if both students are enrolled in the same college physics course and are given a pretest such as the FCI, we may expect the student who took high-school physics course to have a higher score on the physics pretest. On the other hand, Eq. (7) predicts that the two students would achieve similar normalized gains under the assumptions that (a) $\alpha$ depends on a student's learning ability and the effectiveness of the teaching, and (b) $\alpha$ is unrelated to a student's existing content knowledge in a specific area. In this case if we correlate students' general abilities (to be measured through other methods) with their pretest scores of a specific content area, we could expect a very low correlation.

However, it is also possible that when a group of students of similar background but different learning abilities have all been trained in a particular content area. In this situation, their pretest scores in a subsequent course in that area will correlate with both their general ability and the normalized gains in the subsequent course. Therefore, when facing a diverse population, the result really depends on a range of factors such as the compositions of the population, the features of the training of the related content areas among the different subgroups of the population, and how such training was valued by teachers and students and used in the determination of students' performance or ability levels, etc. Since most of these factors are difficult to be precisely determined and controlled with diversely formed population groups, the relation between students' general abilities and their pretest scores in a specific content area that are not consistently trained among all students over a long period of time is often in an undetermined state. That is, depending on the population one may sometimes observe a significant correlation and in some other times and or with different populations the correlation may disappear. Since both situations are possible in reality, the correlation measure becomes more of an empirical question that reflects the actual features of the population.

Then if one collects pretest data from a large number of very diverse population groups that have received different training in the content domain, the pretest results will contain large contributions from the training and other contextual factors (which are independent of the students' learning ability), and therefore will have a weak correlation to students' learning ability (represented with $\alpha$). Based on Eq. (7), students' normalized gains are dependent on $\alpha$ only. If $\alpha$ is not correlated with the pretest score, the normalized gain is then not correlated with the pretest score. The analysis discussed above is consistent with the Hake's results and is also consistent with the observed non-trivial correlations in many studies based on small numbers of population groups. *Note that the unique feature in Hake's results is the large number of different types of populations (that might have received different training) rather than the sizes of individual population groups.*

Mathematically, we can assume that a student's pretest score ($s_{k0}$) in a specific domain is in general related to the student's general ability of learning ($\alpha_k$) and his/her experienced training ($\widetilde{T}_k$) in the corresponding content domain. Since the training in a content domain can exist in many forms and with different contexts, the variable representing this training should be generally in a complicated form, which will not be explored in this study. Based on these assumptions, we can write

$$s_{k0} = f(\alpha_k, \widetilde{T}_k). \tag{8}$$

For a give group of students, there will be a spectrum for each of $\alpha_k$ and $\widetilde{T}_k$. The variance in $s_{k0}$ is the result of the variances in both $\alpha_k$ and $\widetilde{T}_k$. For students at certain age and or education level, one can often assume a somewhat stable range of learning ability. If the group of students have had identical training in a content domain, the variation in $\widetilde{T}_k$ among different students will be small and the variance in $s_{k0}$ will be largely accounted for by the variance in $\alpha_k$. In this case, one would expect a large correlation between $s_{k0}$ and $\alpha_k$, which will then result in a considerable correlation between the normalized gain and the pretest score.

On the other hand, if the group of students all have very different background (or a large population is made of many diverse subgroups), their experience in training on a content domain can be very different. This often lead to a large variance in $\widetilde{T}_k$, while the variance in $\alpha_k$ can still be in the normal range since the students are in the same age level. In this case, the variance in $s_{k0}$ will contain a significant to even a dominant portion due to the variance in $\widetilde{T}_k$. Since $\alpha_k$ and $\widetilde{T}_k$ are independent of each other, one should expect a small correlation between $s_{k0}$ and $\alpha_k$, which will lead to a small correlation between the normalized gain and the pretest score.

In summary, when assuming that the learning is dominated by the α-process and that α doesn't correlate with students pretest scores, a low correlation between the individual students' normalized gains and their pretest scores can be expected. The normalized gain is always positively correlated with students' general learning abilities and the effectiveness of specific instructional methods, which need to be measured with additional methods or through controlled studies.

In a recent report, Coletta et al used Lawson's Classroom Test of Scientific Reasoning to measure the students' scientific reasoning ability (a somewhat general ability) and found a significant positive correlation between FCI gains and Lawson test scores [16-18]. In the present model, the α coefficient is defined to parameterize the effects of learning which is highly related to the general abilities of students; therefore, one can interpret the Lawson test results as some type of measure of the α coefficient, which has a positive correlation with the normalized gain as predicted from Eq. (7). Since the α coefficient is (a) positively correlated with the normalized gain as indicated in Eq. (7), and (b) determined by both the effectiveness of the instruction and student learning abilities, and since the latter are probably reflected in the Lawson test score, the results of Coletta et al. are consistent with the present model.

Using Eq. (7), we can calculate $\overline{g}$, the class average of the individual gains:

$$\overline{g} = 1 - \frac{1}{N}\sum_{k=1}^{N} e^{-\alpha_k T} . \tag{9a}$$

Now define $\overline{\alpha}$ as population's average teaching-learning impact coefficient, and write

$$e^{-\overline{\alpha}T} = \frac{1}{N}\sum_{k=1}^{N} e^{-\alpha_k T} . \tag{9b}$$

We then have $\overline{\alpha}T = -\ln(1-\overline{g})$, which provides an estimation of the overall impact of the instruction on a student population. One can further isolate the effects of *T*; but this will certainly introduce additional uncertainties.

Notice that the α-process describes predominantly the impact from external influence on students' measured-wrong knowledge (no internal interactions). Therefore, the normalized gain can be interpreted as a measure to gage the external impact to the learning from the instruction. As discussed earlier, in studies with certain populations it is common for the normalized gain to have some kind of correlation with the pretest score. Such results may suggest certain dependence between the α coefficients and students' pretest scores and or additional processes such as internal interactions between students' correct and incorrect knowledge during learning, which is modeled with the γ-process to be discussed in the next section.

Hake used the class average scores of pre-post tests to calculate the normalized gain. To distinguish it from the individual students' gains, it is referred here as the population gain, denoted with g:

$$g = \frac{\sum_{k=1}^{N}(1-s_{k0})\cdot(1-e^{-\alpha_k T})}{\sum_{k=1}^{N}(1-s_{k0})} . \tag{10}$$

Compared to Eq. (7), the population gain is an average of $g_k$ weighted over the measured wrong knowledge $(1-s_{k0})$ and has an explicit component of the pretest score [19]. However, Hake's analysis was to look at a large number of classes (or population sample groups), each of which can be treated as an individual unit. In this way, a class' gain is equivalent to an individual student's gain defined in Eq. (7). Basically, one can replace the individual student's score variable in Eq. (6) and (7) with the class average score and all the results remain the same but are interpreted as the outcome for a class as one unit. Therefore, Hake's analysis method and results are consistent with the results from Eq. (7).

In education assessment, there are always random noises in the measurement, which can have non-trivial impact to the results. Here, only the non-cognitive based random noise is considered; therefore, there is no change to the assumptions of the learning process and Eq. (1) remains the same. Combining the random measurement errors in both pre and post tests, the solutions are:

$$s_k(0) = 1 - (1-s_{k0}) + \varepsilon_{k0}$$
$$s_k(T) = 1 - (1-s_{k0}) \cdot e^{-\alpha_k T} + \varepsilon_{kT} \qquad (11)$$

where $\varepsilon_{k0}$ and $\varepsilon_{kT}$ represent the random noises in pre and post test scores. $s_k(0)$ is the pretest score obtained in one particular measurement and $s_{k0}$ is the "true" pretest score that can be estimated by making multiple identical measurements.

Here no details of the various possible origins of such noise are considered. The only assumption is that the noise is of random nature and exists in measurement. Assume that the amplitude of the random noise is small compared to $(1-s_{k0})$. One can calculate the first order approximation of $g_k$:

$$g_k = (1 - e^{-\alpha_k T}) - e^{-\alpha_k T} \cdot \frac{\varepsilon_{k0}}{(1-s_{k0})} + \frac{\varepsilon_{kT}}{(1-s_{k0})}. \qquad (12)$$

The correlation between $g_k$ and $s_{k0}$ is no longer zero since both contain the same random noise, $\varepsilon_{k0}$. The first order approximation of the correlation is

$$r(g_k \cdot s_{k0}) = \frac{\overline{[-e^{-\alpha_k T}/(1-s_{k0})]}}{\sigma_g \cdot \sigma_{s0}} \sigma_{e0}^2, \qquad (13)$$

where $\sigma_{s0}$, $\sigma_{e0}$, and $\sigma_g$ are the standard deviations of the pretest score, the random noise of the pretest score, and the individual normalized gain respectively.

Obviously, the random noises in the measurement of pretest scores will always result in a negative correlation, which is evident from Eq. (7) in that the partial derivative of $g_k$ over the pretest score is always negative and that only the random error portion of the pretest scores contributes to the correlation. A rough estimation using Eq. (13) with a typical class suggests that a random noise of 0.1 can contribute to about $-0.2$ in correlation.

## IV. Model of Learning with Associative Processes

The model discussed in the previous section assumes that the students' existing correct knowledge does not contribute significantly to the learning and that the change of the students' knowledge is primarily due to interactions between the instruction and students' incorrect or missing part of knowledge. Suppose the effects of students' existing correct knowledge are also significant. Then both the *α*-process and the *γ*-process need to be considered. The *β*-process is still assumed to be much less likely to occur compared to the other two processes. This leads to

$$\frac{ds_k}{dt} = \alpha_k(1-s_k) + \gamma_k s_k(1-s_k), \tag{14}$$

which gives

$$s_k(t) = 1 - \frac{\alpha_k/\gamma_k + 1}{1 + \frac{\alpha_k/\gamma_k + s_{k0}}{1 - s_{k0}} e^{(\alpha_k + \gamma_k)t}}. \tag{15}$$

One can further rewrite Eq. (15) as

$$e^{-(\alpha_k+\gamma_k)t} = \frac{1 - s_k(t)}{1 - s_{k0}} \cdot \frac{\alpha_k/\gamma_k + s_{k0}}{\alpha_k/\gamma_k + s_k(t)}. \tag{16}$$

In this case, the normalized gain for a student is found to be

$$g_k = \frac{1 - e^{-(\alpha_k+\gamma_k)t}}{1 + \frac{1 - s_{k0}}{\alpha_k/\gamma_k + s_{k0}} e^{-(\alpha_k+\gamma_k)t}}. \tag{17}$$

It is easy to see that when the *γ*-process is not considered ($\gamma_k = 0$), Eq. (17) reduces to Eq. (7), which doesn't contain explicit terms of $s_{k0}$ and the correlation between $g_k$ and $s_{k0}$ becomes zero if $\alpha_k$ is uncorrelated with $s_{k0}$. In this case, the normalized gain gives a measure of the effectiveness of the α-process. When the γ-process is considered, $g_k$ always has explicit terms of $s_{k0}$ and a positive correlation between $g_k$ and $s_{k0}$ can be expected. It can also be shown that the correlations between $g_k$ and all the coefficients (α, β, and γ) are always positive.

On the other hand, if only the γ-process is considered, one can still separate the ability variables with the scores, which is useful in assessing features that are independent of the pretest scores. In this case, Eq. (16) can be simplified into Eq. (18), which provides a measure of the effectiveness of the γ-process.

$$e^{-\gamma_k t} = \frac{1 - s_k(t)}{1 - s_{k0}} \cdot \frac{s_{k0}}{s_k(t)}. \tag{18}$$

The β-process has been ignored in this discussion, assuming that it normally plays an insignificant role in student learning under current education settings. However, it is easy to see that when considered, the β-process will contribute to a positive correlation between students' gains and pretest scores.

The theoretical model discussed above provides an explanatory framework for the experimental results of normalized gain in existing work [e.g., 1c, 16-17, 23]. In addition, it leads to predicted relations among the normalized gain, student pretest score, student general ability, and effectiveness of teaching. (The separation of students' general ability and the effectiveness of teaching is non-explicit in the model but can be achieved with controlled experiment designs, which will be discussed in section V.)

In summary, *under identical instruction if students' general abilities in learning (represented with α and γ coefficients) are uncorrelated with their scores of a pretest on a specific content area, the correlation between individual students' normalized gains and their pretest scores will be zero if only the α-process is considered, and be greater than zero if the γ-process is considered. Any random noise in the pretest scores will result in a negative contribution to the correlation between $g_k$ and $s_{k0}$. Furthermore, in all situations $g_k$ is always positively correlated with students' general abilities in learning represented with the α, β, and γ coefficients.*

The theoretical predictions agree well with an increasing number of experimental studies, which have documented

- a positive correlation between Conceptual Survey in Electricity (CSE) normalized gains and students' math pretest scores (r = 0.1~ 0.4, N ~150 from 3 classes) [24],
- a positive correlation between FCI normalized gains and spatial visualization ability ( r ~ 0.3, N ~ 60) [25],
- a positive correlation between FCI normalized gains and Lawson test scores (r = 0.51 N=65, similar results were found at OSU) [16], and
- a positive correlation between FCI normalized gains and pre-instruction SAT scores at one high school (N = 335, r = 0.57) and one university (N = 292, r = 0.46) [17].

It is expected that as both theoretical and experimental research advances, a more insightful understanding of learning and measurement can be developed.

## V. Numerical Analysis and Model Evaluation

Based on the theoretical models, numerical methods can be developed to estimate parameters about students' learning. For example, with only the $\alpha$-process, students' change scores can be written as

$$\Delta s_\alpha = s(t) - s_0 = (1 - s_0)(1 - e^{-\alpha t}) = (1 - s_0) \cdot g. \tag{19}$$

If we plot the change vs. the pretest score, it will show as a straight line from point $(0, g)$ to point $(1, 0)$.

Similarly, with only the $\gamma$-process, students' change scores become

$$\Delta s_\gamma = s(t) - s_0 = \frac{1 - e^{-\gamma t}}{1 + \frac{1 - s_0}{s_0} e^{-\gamma t}} \cdot (1 - s_0). \tag{20}$$

With both $\alpha$ and $\gamma$ processes, the change score is

$$\Delta s_{\alpha\gamma} = \frac{1 - e^{-(\alpha+\gamma)t}}{1 + \frac{1 - s_0}{\alpha/\gamma + s_0} e^{-(\alpha+\gamma)t}} \cdot (1 - s_0). \tag{21}$$

Figure 3 shows the change score vs. the pretest score relation for both the $\alpha$-process and the $\gamma$-process. The solid straight lines are the $\alpha$-process relations while the dashed curve lines show the $\gamma$-process relations. These lines are computed with $e^{-\alpha t}$ or $e^{-\gamma t}$ equal to 0.1, 0.3, 0.5, 0.7, and 0.9. For any point on the graph, $g$ can be computed with $\Delta s / (1 - s_0)$.

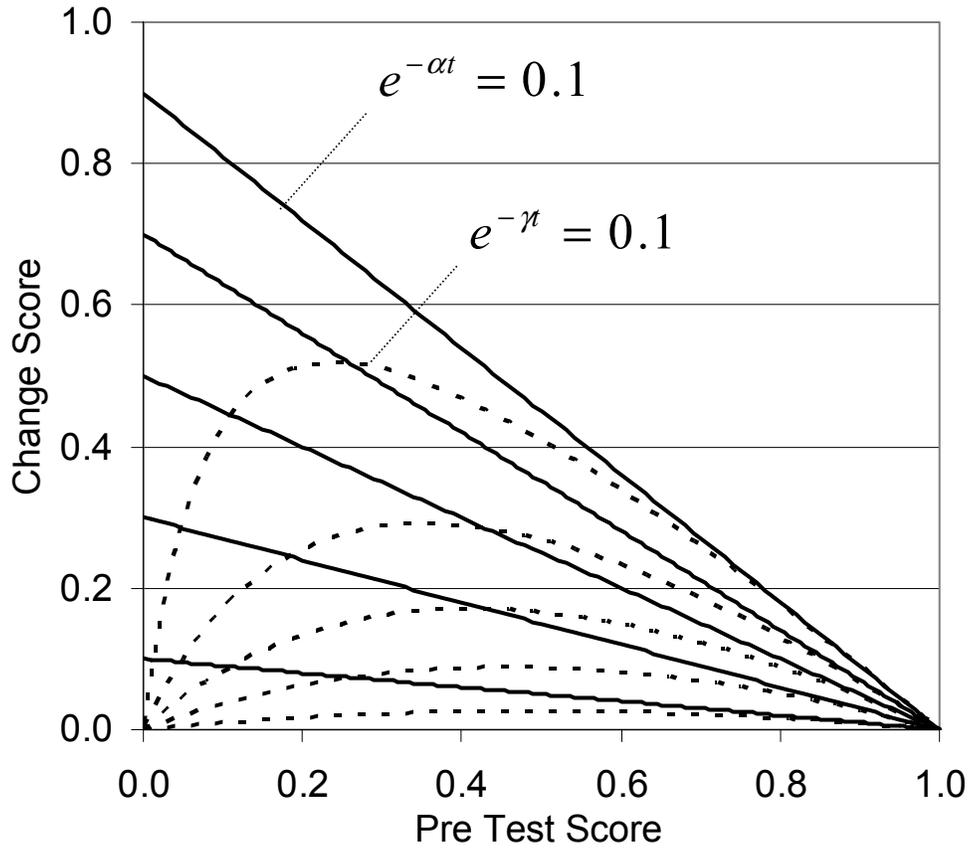

FIG 3. Actual change score vs. pretest score for five values of $e^{-\alpha t}$ or $e^{-\gamma t}$ (0.1, 0.3, 0.5, 0.7 and 0.9) for $\alpha$ (solid lines) and $\gamma$ (dashed lines) processes.

In practice, one can fit the actual data with one or a combination of these models and thereby explore parameters relevant to student learning and evaluate the models. Detailed explorations of data fitting and modeling will be reserved for future discussions. Here a simple example is given to show the general features of such effort. Figure 4 shows the real FCI data from three different calculus-based mechanics courses for science majors: class 1 used interactive labs ($N_1 \sim 350$); class 2 used typical traditional teaching ($N_2 \sim 1500$); class 3 was for honor students ($N_1 \sim 200$) and used traditional lectures with additional emphasis on group work and context-rich problem solving in recitation and homework. The data points shown are binned averages. The standard errors for these data points are less than 0.02.

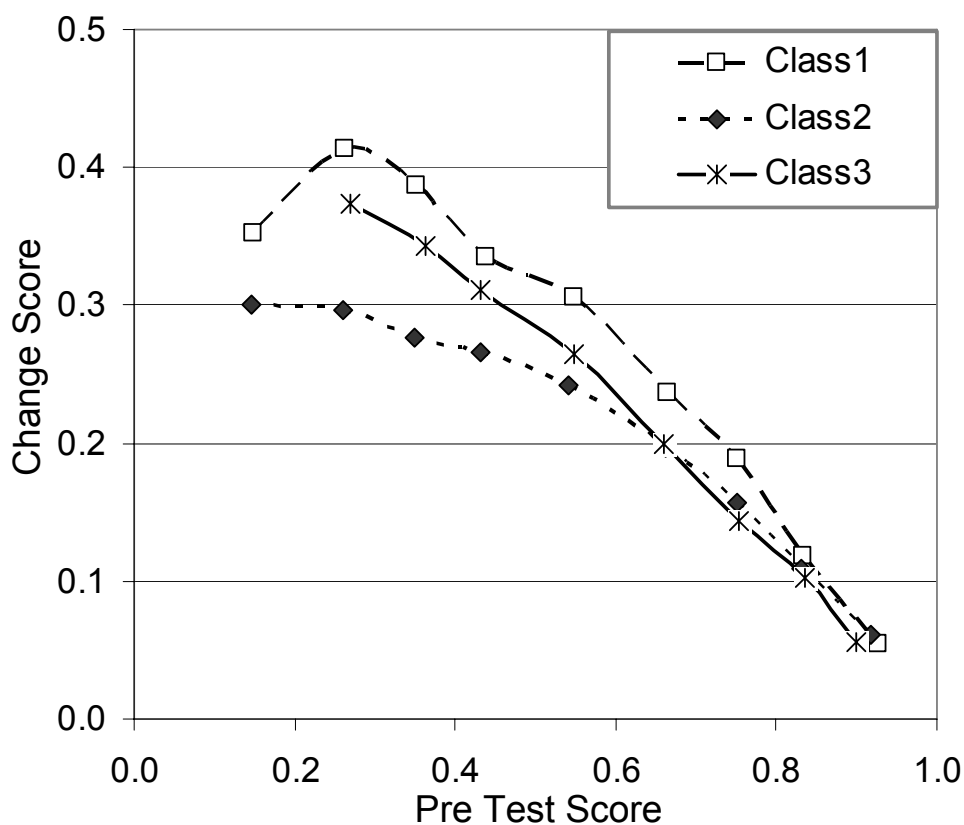

Figure 4. FCI change score vs. pretest score relations of three different calculus-based mechanics classes for science and engineering majors.

A. *Analysis of Possible Learning Processes*

In general, the change score shows a fairly linear relation with the pretest score especially for the part where pretest score is greater than 0.5. This is consistent with the theoretical curves shown in Figure 3, in which the logistic model ($\gamma$-process) shows more non-linear trend with pretest score below 0.5. From the data, one can see that Class 2 has a more apparent "curving down" trend at low pretest score indicating more non-$\alpha$ processes and or non-constant $\alpha$ in those regions. Based on the models, there can be at least two types of mechanisms accounting for this data pattern.

One is that students with low pretest score might possess deeply rooted "misconceptions" which can significantly affect their learning of the correct knowledge. As shown from research, students' incorrect answers to FCI in pretests are often not the results of guessing but rather the answers concentrate on traits of a common set of "misconceptions" [3-7]. It is commonly agreed that students can develop well-established alternative conceptions before and or during instruction, which can compete with the development of correct understandings and can act as barriers to their learning of correct knowledge. Since these students have to constantly reconcile between correct and incorrect conceptions, their learning will show more associative forms ($\gamma$ processes), in which students' existing incorrect knowledge can slow down the learning process and may even result in negative contributions to the outcome of a learning incident. Therefore, at low pretest scores, the learning process can have a significant portion of the $\gamma$-process (or a smaller $\alpha$ for variable $\alpha$ models, see discussions below), which is consistent in part with the data pattern in Figures 3 and 4. On the other hand, if students have already an established somewhat correct knowledge frame, the learning will be simply incorporating new knowledge pieces into the existing knowledge frame and thus show more of the $\alpha$-process. Likewise, if the students are "quick" learners who can restructure their incorrect knowledge frame easily, the learning will also be less impacted by the students' incorrect knowledge and show more of the $\alpha$-process. These extrapolations are consistent with the data patterns of students with high pretest scores and or high learning abilities. Here, the learning ability can be evaluated with information other than the test data such as the populations' background, e.g., honor students vs. average students.

The impact of the incorrect knowledge is also related to the nature of the learning of a scientific concept, in which the learner often has to restructure the existing knowledge frame so that it is consistent across the knowledge domain [8]. One may predict that if the learning is about a pure memorization task, in which students' existing knowledge is less likely to interact with the learning of new knowledge pieces, then learning will behave mostly as the $\alpha$-process even at the low pretest score region.

For students with medium to high pretest scores, they are in the "mixed states" of understanding; therefore, their incorrect knowledge are less stable and easier to change than the students in a pure incorrect state [3,4,12]. These students also have fewer incorrect pieces to battle with so that the "negative drag" on learning due to the incorrect knowledge is also less. Therefore, these students' learning processes will have more $\alpha$-type than the students with low pretest scores. These are consistent with the more linear data pattern at the higher end of the pretest scores.

Another possible explanation is that students' learning ability represented in part with the $\alpha$ coefficients might have a positive correlation with the pretest scores, which can be defined as a variable $α$ model. This doesn't necessarily mean that the $α$ coefficient has a direct causal relation with the pretest scores one way or the other; but rather that it is possible that under certain education settings and for certain populations the two variables might have interactions either directly or with a third variable (or a group of additional variables) and thus develop some kind of systematic relations. In such cases, the normalized gain will also have a positive correlation with the pretest score.

For example, the Lawson Test measures some form of scientific reasoning ability which can be related to the types of ability modeled with the α coefficient [20]. As predicted by the model, if the α coefficient has a correlation with the pretest score of a specific content area, the normalized gain will also have a correlation with the pretest score. Therefore, the Lawson Test score may correlate (can be at different levels) with both the normalized gains and the pretest score [16]. However, it has also been observed in our research with many different population groups that this type of result is highly context dependent in terms of education settings and features of student populations.

In summary, the relation between change score and pretest score can provide implications about features of the learning process. A linear relation between change score and pretest score often indicates a predominant $α$-process with constant $α$ and the correlation between normalized gain and pretest score will be zero. If the data show curving down at low pretest score region, it will give a positive correlation between normalized gain and pretest score and suggest the involvement of $γ$ type processes and or non-constant $α$ with $α$ being smaller in the low pretest score region.

B. *Comparing High and Low Ability Students*

In Figure 4, the data pattern of the honor students (Class 3) shows a more straight linear relation indicating a dominant $α$-process even at low pretest scores. Several independent studies with students at MIT and Harvard have also shown similar linear relations at the low pretest score region [16]. Here we operationally define high ability students to be those from top universities and or in honor programs of an average university. A possible explanation for this phenomenon is that when students' ability and or the effectiveness of the instruction are high, the impact from students' incorrect knowledge on their learning of correct knowledge is less. These students are more likely to be able to restructure their incorrect knowledge framework and

*22*

develop the correct understanding in given learning incidents that address those unknown or incorrect knowledge areas, producing a somewhat constant average success rate of learning at for students at both low and high pretest score regions. In other words, the amount of incorrect knowledge and the pretest score is unrelated to the students' ability to learn. Therefore, learning appears largely in the *α*-process.

There may also be a threshold effect: when students' ability and or the amount of correct knowledge is above certain level, their learning is less affected by their initial incorrect knowledge. Obviously, this effect is dependent on the context of the learning such as the content difficulty.

In the study by Coletta et al (2005), four populations all using interactive engagement instruction were studied [16]. Except for the Harvard students ($r = 0..04$), among the rest three populations which were from typical average universities, a positive correlation between individual students' normalized FCI gains and their pretest FCI scores was reported ($r = 0.15 \sim 0.33$). This study can be considered a typical control variable design of different populations going through similar instruction and the result is consistent with the current model and explanations. The Harvard students are of higher ability compared to students in average universities and their learning behaves in *α* type processes, while the learning of students from average universities involves more *γ* type processes.

A numerical method to study the possible mechanisms and features of the learning processes is to fit the experimental data with the theoretical learning models and explore the features of the computed coefficients. For example, to explore the features of possible *α* and *γ* processes with experimental data, one can use Eq. (21) to fit the data. Rewrite Eq. (21), we have:

$$\Delta s_{\alpha\gamma} = \frac{1 - e^{-(\alpha+\gamma)t}}{1 + \frac{1 - s_0}{\alpha/\gamma + s_0} e^{-(\alpha+\gamma)t}} \cdot (1 - s_0) = \frac{1 - \eta}{1 + \frac{1 - s_0}{\xi + s_0} \eta} \cdot (1 - s_0), \tag{22}$$

where

$$\eta = e^{-(\alpha+\gamma)t}$$
$$\xi = \frac{\alpha}{\gamma} \tag{23}$$

The results of fitting with data from classes 2 and 3 are shown in Figure 5. The *α*-to-*γ* ratio is 1.55 in class 2 and 0.30 in class 3, indicating a dominant *α*-process for class 2 and a strong *γ*- process for class 3. Since class 2 is the honor students, this result is consistent with both the theoretical explanations and the experiment results of the Harvard students reported in Coletta et al (2005) [16].

In both classes, $\eta \approx 0.39$, indicating a similar learning effectiveness at the high pretest score region ($\Delta s$ depends primarily on $\eta$ when pretest score is large). Since class 3 has a significant portion of the $\gamma$-process; therefore, the normalized gain will be positively correlated with the pretest score. Based on limited data sets, it has been observed that for populations similar to Class 3 students, the correlation between the individual students' normalized gains and the pretest scores is around 0.3; while this correlation is less than 0.1 for populations similar to Class 2 students.

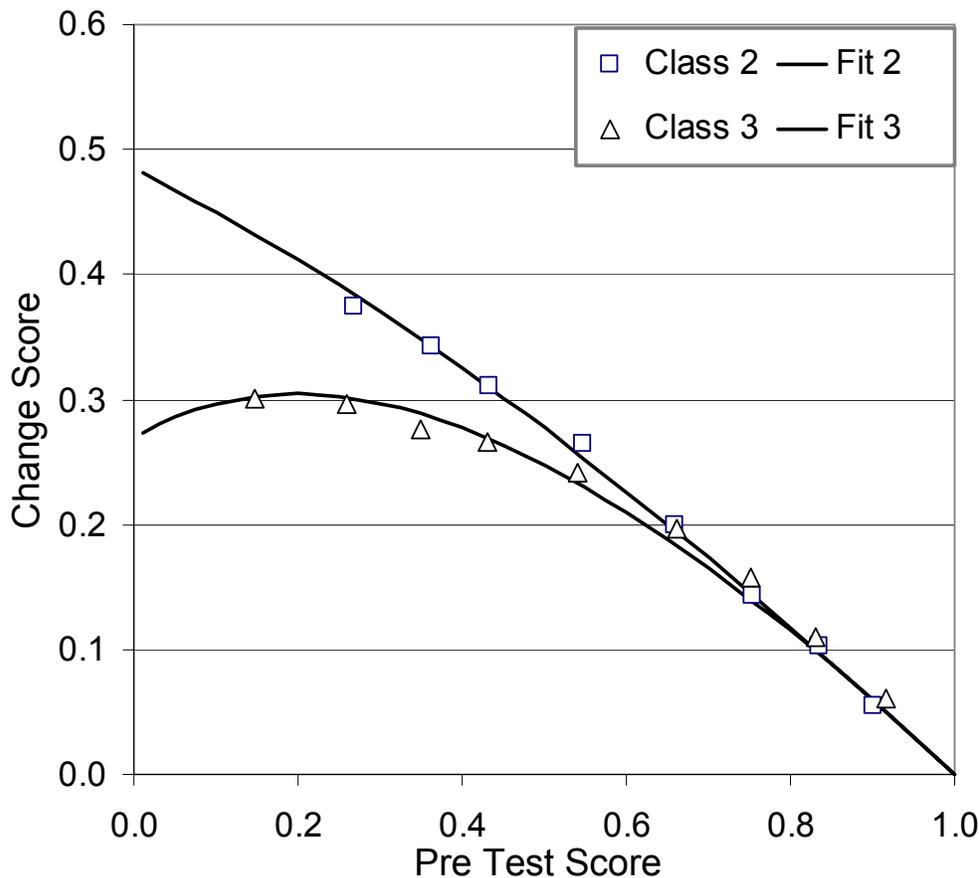

Figure 5. Fitting the change score with both the $\alpha$-process and the $\gamma$-process.

C. *Distinguishing Student Ability and Instruction Impact*

The model coefficients ($\alpha$ and $\gamma$) represent the effectiveness of teaching-learning interactions; therefore, these are related to both students' innate learning abilities and the features of the instruction. With a pre-post measurement on certain content areas over a period of learning, the learning outcome comes from a combined effect of students' abilities and instruction, which cannot be further distinguished. However, by using additional measures and controls such as measurement of students' reasoning ability and or measurement of students' long term course grades (or GPA), one can operationally control the two variables and make assessment of their individual effects.

For example, based on students' background and long term accumulated academic achievement, one can often categorize students into performance groups. The discussion in the previous section about high and low ability students is one case of such variable control, in which the instructional format was held constant and students' ability, operationally defined by students' status such as average students, honor students and MIT students, was assumed to be different. This allows the analysis of the learning behaviors of different populations going through the same style of instruction. On the other hand, one can also compare groups of the same population going through different instruction, which will provide information about the effectiveness of the instructional methods. These are all typical control-variable methods used in education research.

Using this model, the experimental results can be further analyzed in a new perspective. When the instruction method is the same, high ability students will show more of the $\alpha$-process and low ability students will behave more like the $\gamma$-process. (Note that the student ability has to be determined with methods other than the pre or post test scores.) When the population is the same, more effective instructional method can make the students behave more towards the $\alpha$-process and also make the slope of the declining curve in Figure 5 more negative (larger normalized gains). It has to be noted that since the teaching learning interactions are processes of a very complex system, the different types of predictions can have many variations when the impacts of the instruction on different subgroups (e.g. low and high ability students) are different and or when the pre-instruction knowledge (pretest scores) distributions of different subgroups vary. Therefore, variable control is crucial and extensive experimental results with a large number of diverse population groups are needed in order to single out the effects of a specific variable.

However, analysis of limited cases can still provide interesting insights. For example, suppose one compares two instruction treatments with a single population. Treatment 1 produces a $\gamma$-process and treatment 2 also produces a $\gamma$-process but with a more negative slope at high pretest score region. This shows that treatment 2 actually helps students with high pretest scores more than students with low pretest scores. On the other hand, if treatment 2 moves the students towards a $\alpha$-process, the results then suggest a more uniform and favorable impact on students with both high and low pretest scores.

Understanding and distinguishing features of population and instruction are often the core issue in education measurement. With this new model, researchers can be equipped with more tools to work on the problem. For example, one can compare if the learning is more of a $\alpha$ or a $\gamma$ process, how much the slope varies, whether or not there is significant sub-population-group effects and if so one may further conduct data fitting for the sub populations, which will lead to a whole set of new methods and tools. For instance, one may assume different $\alpha$ to fit different subgroups of a population and may also fit the data with a weighted linear combination of the $\alpha$ and $\gamma$ processes. One advantage of this model over the assumptions in IRT is that the parameters assumed are directly related to the features of the learning processes and by controlling variables and conducting additional measures one can make real measurement of how these parameters vary with conditions in the actual learning settings and their impacts on learning outcomes. This provides the capability to adaptively revise and validate the theoretical model so that it can gradually improve its explanatory power of reality.

**VI. Further Generalizations**

The model discussed in this paper suggests a variety of new areas for both theoretical and experimental studies on assessment. A few such examples are summarized below, the details of which will be reported in future papers.

A. *Generalized Forms of Learning Dynamics*

More generally, one can consider the $\gamma$-process a special case of the $\alpha$-process with the $\alpha$ being linearly dependent on the score (non-constant $\alpha$ model). In theory, $\alpha$ can be a function of the score in a more complex form including constant, linear, and higher order terms:

$$\frac{ds}{dt} = \alpha(s) \cdot (1-s) + C, \tag{24}$$

where

$$\alpha(s) = \alpha_0 + \alpha_1 s + \alpha_2 s^2 + \ldots. \tag{25}$$

The existence of any non-constant terms in $\alpha$ will result in a positive correlation between the normalized gain and the pretest score and a "curved down" pattern at low pretest score region in the data plot of change score vs. pretest score.

The meaning of Eq. (24) reflects a measurement based probability model that would lead to a general form "error reduction" process, in which the amount of measured incorrect knowledge is reduced (if $\alpha$ is positive) as a result of learning. The impact of students' measured correct knowledge is included as substructures in the general form of $\alpha$ coefficient.

B. *Connections to Rescorla-Wagner Model and Error-Controlled Learning*

In the field of psychology, Rescorla-Wagner model [21] was developed in the early 70's to describe the changes in associative strength between a signal (conditioned stimulus, CS) and the subsequent stimulus (unconditioned stimulus, US) as a result of a conditioning trial. The model states that the change of the association strength is proportional to the difference between some maximum limits of learning and current associative strength:

$$\Delta V = \kappa \cdot (\lambda - V). \tag{26}$$

Here, $V$ is the current association strength and $\Delta V$ is the change of association strength resulted from a learning trial. $\kappa$ is a constant representing features of the learner and the settings of the learning. $\lambda$ represents the maximum association strength that learning can produce.

The mathematical form of the Rescorla-Wagner model and the $\alpha$-process are identical. However, the Rescorla-Wagner model is about the actual microscopic learning rule for learning a particular set of associations, which assumes that error is used by the learning mechanism to determine how the association strength should change. This type of error –controlled learning is the basis for many standard regression methods and popular learning algorithms such as back-propagation neural networks (BPN). The result of such learning shows an error-correction or error-reduction type of process.

On the other hand, the dynamic model discussed in this paper is based on a simple probabilistic representation of measured knowledge states and learning incidents. The measured wrong knowledge states are not used by the learning mechanism to change knowledge states. The changes of the measured knowledge states are simply the probabilistic results of learning incidents addressing unknown knowledge, which is a process of random origin.

The fundamental difference between the dynamic model and the Rescorla-Wagner model (or other methods such as BPN) is in the way how error is involved in the learning. In the Rescorla-Wagner model and other error-reduction algorithms, error is used by the learning mechanism to determine and manipulate the actual processes of learning. In the dynamic model presented here, the error (incorrect knowledge) is simply a statistical reality, which doesn't materially control the learning. The error is simply the target for the learning incidents to make impact. However, both the dynamic model and the Rescorla-Wagner model will produce the same type of error-reduction learning curve and mathematical formulism, although the mechanisms leading to the error-reduction process are different within the two models.

In addition, the dynamic model doesn't make assumptions on any particular type of processes that might be happening at the microscopic level. The effects of microscopic learning processes are summed in the coefficients such as $\alpha$ and $\gamma$, the substructures of which are not included in the model. Therefore, this model is rather independent of the actual microscopic learning processes and can be a more general descriptive framework for analyzing macroscopic learning outcomes.

Stretching it further, one may identify substructures of the neural formation processes of the association strength described in the Rescorla-Wagner model, which can then be the results of applying the dynamic model on the probabilistic distributions of the substructures of the neural associations. In this way, the error-controlled learning mechanism can then be viewed as the statistical results of finer substructures and processes. In general, the two types of mechanisms discussed above can be embedded within each other and support each other. The theoretical insights obtained through comparing the different models can provide a more solid and deeper understanding on error-reduction methods in modeling learning.

C. *Comparing the Probability Frameworks of IRT and Normalized Gain*

The Item Response Theory (IRT) is a popular assessment model used in education measurement. The details of this method can be obtained from a vast collection of existing literature [22, 23], which will not be reviewed in this paper. This section highlights the theoretical foundation of IRT and compares its probability framework with that of the normalized gain.

The IRT is designed to evaluate students' performance on certain content areas that are scalable with some ability dimensions. The basic IRT model assumes a unidirectional ability scale and a known (usually normal) distribution of the population along the ability dimension. The probability framework of IRT builds on these assumptions to describe a functional relation between students' ability scale and students' performance measures. This functional relation is used to quantitatively predict the probabilities of students' performance on given tests. This function is then used in nonlinear maximum likelihood regressions to estimate variables that parameterize students' ability and features of test items (such as the item difficulty and discrimination factors).

This non-linear functional relation helps addressing the floor and ceiling effects in assessment. In most cases, the main goal of such measurement is to find students' performance level. The assessment is often conducted with a single cross-sectional measurement of the interested population. The probability framework is about the relation between assumed students' innate abilities and students' performances on given tests. Therefore, in the IRT model, the probability framework, the related measurement methods, and the assessment goals do not explicitly model and address how students' knowledge is changed over time, i.e., no parameters are included in the model to represent the dynamic process of learning and features of the teaching-learning environment.

On the contrary, the normalized gain is designed to measure the effectiveness of students' learning during a short period of time independent of students' initial knowledge states. Therefore, by design the normalized gain is a measure of change that requires multiple measurements (≥2). However, since its wide use from a decade ago, the theoretical basis and the associated probability framework of the normalized gain have not been formally discussed. The model discussed in this paper provides a theoretical explanation of features of the normalized gain and a probability framework for its dynamics. The meaning of the probability framework can be summarized in one sentence: "*Changes of measured knowledge states depend not only on students' ability and or effects of instruction but also on the amount of students' incorrect and unknown knowledge.*" Compared to IRT, the normalized gain method uses a completely different type of probability framework, which emphasizes the teaching-learning environment and interactions such as the probabilistic feature of instruction addressing specific difficulties. This is particularly important to the measurement of changes of knowledge.

Through comparison, it is easy to see that the probability frameworks, goals, and measurement methods of IRT and normalized gain are all different. The IRT assumes a category of students' innate abilities and their functional relations to students' performances on tests, based on which IRT methods can estimate parameters representing students' abilities as well as features of the test items with one-time measurement data. On the other hand, the normalized gain is a measurement of change, which depends on students' ability, the effects of instruction, and the amount of students' incorrect and or missing knowledge.

Seeing the differences is the first step for developing a more fundamental understanding of education measurement. The learning dynamic model discussed in this paper opens up a new venue of experimental and theoretical work that may eventually integrate many of the existing assessment methods under a coherent theoretical frame and help develop new types of assessment methods based on the insights from the theory.

D. *Conceptualizing the Item Response Function in Item Response Theory*

The core function in IRT is the item response function (IRF), which describes the probabilistic relation between students' ability parameters and the predicted probabilities for their performances on given tests. Based on the basic IRT assumptions, the IRF is a normal cumulative distribution function or the normal ogive, an S-shaped curve that is often approximated with the logistic (or sigmoid) curve for computation simplicity. In Eq. (27), the two-parameter logistic model IRF for dichotomous responses (i.e. either 1 or 0) is given

$$P(\theta) = \frac{1}{1+e^{-D(\theta-b)}}. \tag{27}$$

Here, $\theta$ is a scalar parameter representing a student's ability. $b$ gives the difficulty of a test item, which is in the same scale and dimension with the student's ability. $D$ is the discriminative factor, which controls how steeply the S-shaped curve rises around $\theta = b$. $P(\theta)$ is the probability for a student with ability $\theta$ to give a correct answer to a question with item difficulty $b$ and discrimination $D$.

The S-shaped response curve helps to address the ceiling and flooring effects making the estimation of students' ability parameters be converged in those regions. However, to obtain a deeper insight of the conceptual basis underlying the logarithmic function form, it is useful to further explore the possible connections between basic mechanisms of actual learning processes and the response curve in IRT. The dynamic model in this paper can provide a way to explore such connections. For example, when learning is considered in an associative form, only the $\gamma$- process is kept and Eq. (15) becomes the logistic equation:

$$s_k(t) = \frac{1}{1+(1-s_{k0})/s_{k0} \cdot e^{-\gamma_k t}} = \frac{1}{1+e^{-(\gamma_k t + C_k)}}, \tag{28}$$

where

$$C_k = \ln\left(\frac{1-s_{k0}}{s_{k0}}\right). \tag{29}$$

By comparing Eq. (27) with Eq. (28), one can see a connection between the IRT parameters and the coefficients in the dynamic model:

$$\gamma_k t + C_k = D(\theta_k - b). \tag{30}$$

It suggests that the difference between individual's ability parameter and the item difficult modulated by the discrimination factor (the right hand side of the equation) is equivalent to the accumulated combined effects of the student's learning ability and the instruction impact (the left hand side of the equation). Here, $C_k$ gives the accumulated effects of student ability and learning impact due to prior learning. The physical meaning of the parameters in Eq. (30) is straightforward and provides an explicit link between dynamic changes of students' abilities and students' accumulated abilities in producing correct answers. This is a potentially useful insight that can help the development of both models.

In summary, IRT and the dynamic model each assumes a different set of parameters. The IRT builds on a probability framework for the relation between students' ability parameters and their performances on given test items. The dynamic model builds on a set of explicitly stated learning processes. However, the two different models can produce response functions of similar logarithmic forms. Further, in both models, the assumed parameters are all related to some hypothetical student ability parameters. Therefore, the commonality among the response functions provides a mechanistic justification to relate the space of students' ability to the space of observed probability of students' performance with some forms of logarithmic (or exponential) functions. This understanding is a theoretical advancement for interpreting results from existing models and can help future theoretical development.

E. *The Probability Interpretation of the Normalized Gain*

From a pure probability perspective, the normalized gain can also be viewed as a conditional probability in that it quantifies the likelihood of positive learning outcome under the condition that a useful learning incident is encountered. A useful learning incident is operationally defined as one that addresses a content area that a student either has incorrect or unknown knowledge. To formally model these probabilities, one can make the following definitions.

- Define P(A) as the overall probability for a student to have positive learning outcome independent of learning incidents. This probability can also be interpreted as the student's general learning ability.
- Define P(B) as the probability for a student to encounter a useful learning incident, which addresses a measured-wrong content topic. At pretest, P(B) is given by $P(B) = (1 - s_{k0})$.
- Define P(A|B) as the conditional probability for a student to have positive learning outcome (positive score change) in a useful learning incident.
- Define P(A∩B) as the joint probability for a student to encounter a useful learning incident and to produce positive learning outcome. With pre-post measurement, this probability is equal to the actual pre-post change score when score is scaled within 0 to 1.

Then, using the conditional probability formulation, we can write

$$P(A|B) = \frac{P(A \cap B)}{P(B)} = \frac{s_k(T) - s_{k0}}{1 - s_{k0}} = g_k . \tag{31}$$

Therefore the normalization part in computing the normalized gain is in fact producing a conditional probability that describes the likelihood for a student to have positive learning outcome in a given useful learning incident. This normalization makes this conditional probability re-scaled based on the amount of students' incorrect or unknown knowledge.

If P(A) and P(B) are independent with each other, we have

$$P(A \cap B) = P(A)P(B) \text{ and } P(A|B) = P(A) = g_k \tag{32}$$

This indicates that when students' general learning abilities are independent of their content knowledge in specific areas, the normalized gain gives the measure of a student's general learning ability.

F. *Revisiting the Error-Reduction Interpretation of the Normalized Gain*

The *α*-process can be viewed as a simple direct error-reduction model, in which the instruction interacts only with the students' measured wrong knowledge (the error part) and reduces the amount of it. Built on this interpretation, one can assess the effectiveness of teaching and learning by calculating the percentage of how much "error" has been corrected after instruction, which can be written as

$$\Delta err\% = \frac{(1-x)-(1-y)}{1-x} = \frac{y-x}{1-x} = g. \tag{33}$$

Here, $x$ is the pretest score and $y$ is the post test score. We can see that if the learning is very effective, there will be little error remained after instruction and $\Delta err\%$ will be close to 1. When the learning is ineffective, the error won't change much and $\Delta err\%$ will be close to 0.

Based on Eq. (33), the normalized gain can be seen as the percentage of how much "error" has been corrected due to instruction, which fits nicely with the simple error-reduction model. Therefore, when students' correct knowledge doesn't interact with learning significantly, the effectiveness of teaching and learning can be evaluated by considering only the percentage amount of error reduced by instruction and the students' initial state of measured knowledge becomes irrelevant. Obviously, if students' learning is dependent on their correct knowledge, the normalized gain will become less accurate in assessing learning.

**VII. Discussions**

Quantitative measurement is a foundation area for education research and practice. A good understanding of the measurement models and methodology is of great importance to all areas in education. In recent decades, many popular measurement methods and theories have been developed, which have significantly advanced the field of education and provided powerful tools for educators and researchers to tackle difficult problems. However, at current stage, we have yet to achieve a more fundamental and unified understanding of the mechanisms underlying features of the teaching-learning interactions in education settings and the results of quantitative assessment. The models and methods discussed in this paper is the beginning part of a systematic effort to address this problem.

One significant finding from Hake's study is that the normalized gain shows little correlation with students' pre-test scores, when measured with a large number of different population groups [1c]. The model discussed in this paper can be used to explain this result in terms of three processes: (1) When the learning occurs in a pure $\alpha$ process with constant $\alpha$, the correlation is zero. (2) The random uncertainty in the measurement of pretest scores will always result in a negative component in the correlation. (3) When learning behaves in a more associative form and or involves more interaction with the correct knowledge, a positive component in the correlation is expected. The overall correlation consists of all three components.

Under different education and measurement settings, individual studies often find sizable correlations between students' normalized gains and their pretest scores. This can also be explained with the model in terms of similarities among population samples with respect to prior learning. The work reported by Coletta and Phillips (2005) is a typical example of a control-variable study of different populations under similar instruction. They showed, that FCI normalized gains of students in mainstream university courses were positively correlated with their pretest scores, but that for Harvard students, all with presumably relatively higher learning abilities, the correlation was quite weak. In terms of the present model, the Coletta & Phillips data suggest that the learning of students in mainstream universities may involve significant $\gamma$ type processes, while the learning of Harvard students may be predominantly of the $\alpha$ type with little dependence of $\alpha$ on pretest scores.

With additional measures such as measurement of student ability, measurement of students' math performance [24], scientific reasoning performance and SAT score [16, 17], and spatial visualization ability [25] , researchers have reported significant correlations between students' normalized gains on physics conceptual tests and those measures believed to be related to students' more general abilities of learning. These abilities are related to but not dependent on the specific content areas measured with the normalized gains. These results can also be explained with the present model, which shows that the normalized gain is always correlated with the $\alpha$ and $\gamma$ coefficients. These coefficients are indeed parameters defined to model students' learning abilities.

This new model also explains how the normalized gain addresses the ceiling effect of the measurement. When absolute scores and score differences are used in evaluating students' performance, the result is highly dependent on how the measurement window is structured over the depth and breadth of the entire content area. On the other hand, the present model provides a different approach. It separates the general probability for someone to have a change in score from how much that person's score can be changed. Obviously, the product of the two will produce the actual score change. The advantage of this separation is exactly the goal of the normalized gain – to measure the effects of teaching-learning (or the probabilities for students to change their scores under certain education settings) independent of the students' absolute scores (or how much the students can be improved within the measurement window). For example, a student with a high pretest score will have little to improve within the measurement window. However, this student can still have a high probability of improving his/her score when a unknown content topic is presented, since the actual score improvement is the product of the conditional probability of favorable learning in a useful learning incident and the amount of knowledge pieces to be learned. If absolute change scores are used in evaluating the student's performance, the results of students with high pretest scores will be significantly affected by the ceiling effect. With the separation of the probability, the ceiling effect will have much less impact, since a student can still obtain a high normalized gain even with little to learn.

It has to be emphasized that learning gains are always due to the combined effect of students' ability and the effectiveness of instruction. Actual interpretations of the results depend on how these two variables are controlled.

From a more theoretical perspective, this new model establishes a probability framework for the normalized gain method, which makes it comparable with other more theoretically based methods such as IRT. Further the model suggests that the normalized gain emphasizes a different type of probability framework that is not included in the IRT. This provides new insights for understanding the existing measurement methods and also new directions for further development of measurement theory and methodology. Finally, this model connects well with the fundamental assumptions in IRT and the Rescorla-Wagner model in psychology research; therefore, it provides a basis for bridging education assessment with basic studies in psychology and cognitive research.

In summary, the proposed model supports the empirical results that with controlled populations the normalized gain can be a measurement of the instructional impact. The model provides a theoretical explanation of why normalized gain works in this way. The model also predicts explicit relations among assumptions of the learning process and the measured correlation between the normalized gain and the pretest scores. Methodologies to further analyze the correlation can be derived. Further theoretical analysis suggests that this model can provide a more consistent framework connecting many different theories and models in education and cognitive research.

## V. Acknowledgement

The author would like to thank Prof. E. F. Redish, whose guidance and comments motivated many of the ideas. The author greatly appreciates the formative comments and discussions from members of the OSU Physics Education Research Group throughout the 7-year process of this work, in particular G. Aubrecht, L. Jossem, A. Heckler, and N. Reay. The author also thanks Prof. D. E. Pritchard for his constructive discussions and collaborations on extensions of this work.